\documentclass[submission,copyright,creativecommons]{eptcs}
\providecommand{\event}{FOCLASA 2011} 
\usepackage{stmaryrd,amsmath,amsfonts,amstext,amssymb}
\usepackage{graphicx}
\usepackage{relsize}
\usepackage{xspace}
\usepackage{epsfig}
\usepackage{url}
\usepackage{dsfont}
\usepackage{vaucanson-g}
\usepackage[english]{babel}
\usepackage{array}
\usepackage{MnSymbol}

\newcommand{\n}{\ensuremath{\mathsf{n}}}
\newcommand{\fn}{\ensuremath{\mathsf{fn}}}
\newcommand{\bn}{\ensuremath{\mathsf{bn}}}

\newcommand{\inter}[1]{\stackrel{\scriptsize #1}{\rightarrow}}
\newcommand{\irule}[2]
       {\mkern-2mu\displaystyle\frac{#1}{\vphantom{,}#2}\mkern-2mu}

\newcommand{\NR}{\ensuremath{\mathsf{N}}}
\newcommand{\W}{\ensuremath{\mathsf{W}}}
\newcommand{\SD}{\ensuremath{\mathsf{S}}}
\newcommand{\E}{\ensuremath{\mathsf{E}}}
\newtheorem{fact}{Fact}[section]

\newtheorem{theorem}[fact]{Theorem}

\newtheorem{example}[fact]{Example}
\newtheorem{definition}[fact]{Definition}

\title{Predicting global usages of resources\\
 endowed with local policies
\thanks{Research supported by the Italian PRIN Project ``SOFT'', FET Project ``ASCENS'' and Autonomous Region of
Sardinia Project ``TESLA''.}
}
\author{Chiara Bodei, 
Viet Dung Dinh 
and
Gian Luigi Ferrari
\institute{Dipartimento di Informatica,  Universit\`a di Pisa, Italy}
\email{\{chiara,dinh,giangi\}@di.unipi.it}
}
\def\titlerunning{Predicting global usages of resources endowed with local policies}
\def\authorrunning{C. Bodei, V.D. Dinh, \& G.L. Ferrari}
\begin{document}
\maketitle

\begin{abstract}
The effective usages of computational resources are a primary concern of up-to-date distributed applications. In this paper, we present a methodology to reason about resource usages (acquisition, release, revision, ...), and therefore the proposed approach enables to predict \textit{bad} usages of resources. 
Keeping in mind the interplay between local and global information occurring in the application-resource interactions, 
we model resources as entities with local policies and global properties governing the overall interactions. Formally, our model takes the shape of an extension of $ \pi $-calculus with primitives to manage resources. We develop a Control Flow Analysis computing a static approximation of process behaviour and therefore of the resource usages.

\end{abstract}

\section{Introduction}
Evolutionary programming paradigms for distributed systems changed the way computational resources are integrated into applications. Resources are usually geographically distributed and have their own states, costs and access mechanisms. Moreover, resources are not created nor destroyed by applications, but directly acquired on-the-fly when needed from suitable resource rental services. Clearly, resource acquisition is subject to availability and requires the agreement between client requirements and service guarantees (Service Level Agreement -- SLA). 
The dynamic acquisition of resources increases the complexity of software since the capability of adapting behaviour strictly depends on resource availability.
\emph{Ubiquitous computing} \cite{Abowd2000} and
\emph{Cloud computing} \cite{Buyya2009,you08,EECS-2009-28} provide illustrative examples of a new generation of applications where resource awareness has been a major concern.

The design of suitable mechanisms to control the distributed acquisition and ownership of computational resources is therefore a great challenge. Understanding the foundations of the distributed management of resources could support state-of-the-art advances of programming language constructs, algorithms and reasoning techniques for resource-aware programming. In the last few years, the problem of providing the mathematical basis for the mechanisms that support resource acquisition and usage has been tackled by several authors (see e.g.
\cite{BDFZ09,BuscemiM07,DBLP:journals/lmcs/KobayashiSW06,DBLP:journals/fac/CollinsonP10a,FVH}, to cite only a few).

Here we consider a programming model where processes and resources are distinguished entities. Resources are computational entities having their own life-cycle.
Resources can range from computational infrastructures, storage and data services to special-purpose devices. Processes dynamically acquire the required resources  when available, but they cannot create any resource. This simple programming model abstracts the features of several interesting distributed applications. As an example, let us consider a cloud system offering computing resources. The available resources are the CPU units of a given power and processes can only acquire the CPU time, when available, to run some specialised code. Similar considerations apply to storage services, where client processes can only acquire slots of the available storage. In our programming model, the deployed resources can be dynamically \emph{reconfigured} to deal with resource upgrade, resource un-availability, security intrusion and failures. A distinguished feature of our approach is that the reconfiguration steps updating the structure of the available resources are not under the control of client processes. 

In this paper, we introduce the formal basis of our programming model. Specifically, we introduce a process calculus with explicit primitives for the distributed ownerships of resources. In our calculus, resources are not statically granted to processes, but they are dynamically acquired on-the-fly when they are needed. 
%

We start from the $\pi$-calculus~\cite{SW-Book} and we extend it with primitives to represent resources and the operations to acquire and release resources on demand. 
Central to our approach is the identification of an abstract notion of resource. 
In our model,
resources are \emph{stateful} entities available in the network environment where processes live. Specifically, a resource is described through the declaration of its interaction endpoint (the resource name), its \emph{local} state and its \emph{global} properties. Global properties establish and enforce the
SLA to be satisfied by any interaction the resource engages with its client process.
The global interaction properties can be expressed by means of a suitable resource-aware logic in the style of~\cite{BDFZ09}, or contract-based logic as in~\cite{DCGP09,BZ10}.
The interplay between local and global information  occurring in the process-resource interactions motivates the adjective {\em G-Local} given to our extension of the $\pi$-calculus.

Since we build over the $\pi$-calculus, name-passing is the basic 
communication mechanism among processes. 
Beyond exchanging channel names, processes can pass resource names as well. 
Resource acquisition is instead based on a different abstraction. In order to acquire the ownership of a certain resource, a process
issues a suitable request. Such request is routed in the network environment to the resource. The resource is granted only if it is available. In other words the process-resource interaction paradigm adheres to the \emph{publish-subscribe} model: resources act as publishers while processes act as subscribers. Notice that processes issue their requests without being aware of the availability of the resources. 
When they have completed their task on the acquired resource they release it and make it available for new requests. 
The two-stage nature of the
publish-subscribe paradigm relaxes the inter-dependencies among computational components thus achieving a high degree of loose coupling among processes and resources. 
In this sense our model also resembles tuple-based systems \cite{Gelernter}. 
Consequently, our model seems to be particularly suitable 
to manage distributed systems where the set of published resources is subject to frequent changes and dynamic 
reconfigurations.

To summarise, our approach combines the basic features of the  $\pi$-calculus (i.e.~dynamic communication topology of processes via name passing) with the publish-subscribe paradigm for the distributed acquisition of resources. This is our first 
contribution. The interplay between local and global views is also one of the novel features of our proposal.  A second contribution consists in the development of a \emph{Control Flow Analysis} (CFA) for our calculus. The analysis computes a safe approximation of resource usages. Hence, it can be used to statically check whether or not the global properties of resources usages are respected by process interactions. 
In particular, it helps detecting {\em bad usages} of resources, due to policy violations. 
This suggests where are sensible points in the code that need dynamic check in order to avoid policy violations.

\noindent \emph{Related Work.} 
The primitives for resource management make our approach easy to specify a wide range of the resource behaviour of distributed systems such as Cloud Computing and Ubiquitous Computing. We believe that our approach also leverages analysis technique such as CFA and behavioural types. 
A simplified version of the G-Local $\pi$-calculus has been
presented in~\cite{BDF2011}. The work presented here differs in several
ways from the previous one. 
The version of the calculus we considered in this paper is more expressive of the one presented in~\cite{BDF2011} since here processes can pass resource names around. 
This feature was not allowed in~\cite{BDF2011}. 
Also, the management of resource acquisition and release is much more powerful. 

In~\cite{BDFZ09}  an extension of the $\lambda$-calculus is proposed to
statically verify resource usages. Our notion of global usages is inspired
by this work.
The $\pi$-calculus dialect of~\cite{DBLP:journals/lmcs/KobayashiSW06}
provides a general framework for checking resource usages in distributed
systems. In this approach private names are extended to resources,
i.e.~names with a set of traces to define control over resources. Also
resource request and resource release are simulated through communicating
private names and structural rules respectively.
This gives shared semantics of resources, i.e.~several processes can have
a concurrent access to resources (by communicating private names).
In our approach, when a process obtains a resource, it has an exclusive
access to it.
Furthermore, resource entities can be dynamically reconfigured, while this is not
the case in~\cite{DBLP:journals/lmcs/KobayashiSW06}.

In~\cite{DBLP:journals/fac/CollinsonP10a}, resources form a monoid and
the evolution of processes and resources happens in a SCCS style. 
In our approach, resources are independent stateful entities equipped with
their own global interaction usage policy. A dialect of the
$\pi$-calculus,
where resources are abstractly represented via names and can be allocated
or de-allocated
has been introduced in~\cite{FVH}. In this approach reconfigurations steps
are internalized inside processes via the operations for allocating and
de-allocating channels. A type system capturing safe reconfigurations over
channels has been introduced. In our approach resources are more
structured than channels and their reconfiguration steps are not under the
control of processes.
Finally, the work presented in~\cite{DBLP:conf/esop/BuscemiM07} mainly 
focuses on specifying SLA by describing resources as suitable constraints.
Our approach can exploit constraints to express global resource usages as
well.

\section{The G-Local $ \pi $-Calculus}
\paragraph{Syntax}
We consider the monadic version of $\pi$-calculus~\cite{SW-Book}
extended with suitable primitives to declare, access and dispose resources.
The syntax is displayed in Fig.~\ref{fig:syn}.
Here, $ \mathcal{N} $ is a set of channel names (ranged over by $ x,y,z$), 
$ \mathcal{R} $ is a set of resource names (ranged over by $ r,s,t$) 
and
$\mathcal{A}$ is a set of actions (ranged over by $ \alpha,\beta$) 
for running over resources.
We assume that these sets are pairwise disjoint.
From now on, for the sake of simplicity, we often omit the trailing $\textbf{0}$. 



\begin{figure}
{\small
\[
\begin{array}{lrlllrll}
P,P' &::= & & \ \ \hbox{\it processes}
        \ \
       & \pi,\pi' &::= & & \ \ \hbox{\it action prefixes} \\

& & \textbf{0} & \ \ \hbox{empty\ process} 
& & & \tau & \ \ \hbox{internal\ action}\\

& \;\mid\; & \pi.P & \ \ \hbox{prefix\ action} 
& & \;\mid\; & x(w) & \ \ \hbox{free\ input}\\

& \;\mid\; & (\nu{}z)\ P & \ \ \hbox{restriction} 
& & \;\mid\; & \bar{x}w & \ \ \hbox{free\ output}\\


&  \;\mid\; & P + P' &\ \ \hbox{choice} 
& & \;\mid\; & \alpha(r) & \ \ \hbox{access\ action}\\

&  \;\mid\; & P \parallel P' &\ \ \hbox{parallel\ composition} 
& & \;\mid\; & rel(r) & \ \ \hbox{release\ action}\\

& \;\mid\; & (r,\varphi,\eta) \lbrace P \rbrace & \ \ \hbox{resource\ joint\ point}
\\

& \;\mid\; & req(s) \lbrace P \rbrace & \ \ \hbox{resource\ request\ point} 
\\

& \;\mid\; & !P & \ \ \hbox{replication}
\\

\end{array}
\]
}
\caption{The syntax of G-Local $ \pi $-calculus.} \label{fig:syn}
\end{figure}

{\normalsize

The input prefix
$ x(w).P $ binds the name $ w $ (either a channel or a resource) within the process $P$,
while the output prefix $\bar{x}w.P $ sends  the name $ w $ along channel $ x 
$ and then continues as $ P $.
Note that resource names can be communicated, however they cannot be used as private names and used as channels.
As usual, input prefixes and restrictions act as bindings. 
The meaning of the remaining operators is standard. 
The notions of names $\n()$, free names $\fn()$, bound names $\bn()$ 
and substitution $\{-/- \}$ are defined as expected.

Our extension introduces resource-aware constructs in the $\pi$-calculus.
The access prefix $\alpha(r)$ models the invocation of the operation $\alpha \in \mathcal{A}$
over the resource bound to the variable $ r $. 
Traces, denoted by $ \eta,\eta' \in \mathcal{A}^* $, are finite sequences of events. 
A usage policy is a set of traces. 
The release prefix $rel(r)$ describes the operation of releasing the ownership of the resource $s$. 
In our programming model, resources are viewed as stateful entities,
equipped with policies constraining their usages.
More precisely, a resource is a triple $ (r,\varphi,\eta) $, where $ 
r \in \mathcal{R} $ is a resource name, $ \varphi \in \Phi $ is the 
associated policy and
$ \eta \in \mathcal{A}^*$ is a state ($\epsilon$ denotes the empty state).
Policies specify the required properties on resource usages.  Policies are usually 
defined by means of a resource-aware logic 
(see~\cite{BDFZ09,BZ10,Caires08,DCGP09}),
while states keep track of the sequence of actions performed on resources, by means of (an 
abstraction of) execution traces. 

For instance, in~\cite{BDFZ09}, the policies are expressed in terms of automata over an infinite alphabet, where automata steps correspond to actions on resources and final states indicate policy violations. 

To cope with resource-awareness, we introduce two primitives 
managing resource boundaries: resource joint point $ (r,\varphi,\eta) 
\lbrace P \rbrace $ and resource request point $ req(r)\lbrace P 
\rbrace $.
Intuitively, process $ P $ when plugged inside the resource boundary $ 
(r,\varphi,\eta)\lbrace P \rbrace $ can fire actions acting over the 
resource $ r $.
The state $\eta$ is updated at each action $\alpha(r)$ according to 
the required policy $\varphi$.
A resource request point $ req(r) \lbrace P \rbrace $ represents a 
process asking for the resource $ r $.
Only if the request is fulfilled, i.e.~the required resource is 
available, the process can enter the required resource boundary and 
can use the resource $ r $,
provided that the policy is satisfied.
Processes of the form
$ (r,\varphi,\eta)\lbrace \textbf{0} \rbrace $ represent available resources.
These processes are idle: they cannot perform any operation. In other 
words, resources can only react to requests. 


\begin{example} To illustrate the main features of the calculus, we consider a small example, 
which describes a workshop with two hammers and one mallet. 
Tools are modelled as resource entities: $hammer$ and $mallet$, with the policy $\varphi_h$ ($\varphi_m$, resp.) that one can only make \textit{hard hit} (\textit{soft hit}, resp.) when using $hammer$ ($mallet$, resp.).
We model workers as a replicated process, whose instantiations take a hammer or a mallet to do jobs, whose chain is described by $Jobs$. 
Job arrivals are modelled as sending/receiving $hammer$  and $mallet$ on the channels $x,y$. 
Furthermore, we assume that there are two types of jobs, \textit{hard} jobs on the channel $x$ and \textit{soft} jobs on the channel $y$, which get done by $hard\_hit$ and $soft\_hit$ actions respectively. 

The initial configuration of the workshop is given below. 
Resources ($hammer$ and $mallet$) have empty traces. 
Note that we have two resources of the same name $hammer$, which corresponds to the number of available hammers in the workshop. 
Intuitively, it means that only two jobs, which use hammers, can be concurrently done. 
We have a sequence of four jobs described by the process $Jobs$.

{
\[
\begin{array}{rl}
Tools ::= & (hammer,\varphi_h,\epsilon)\{ \textbf{0} \} | (hammer,\varphi_h,\epsilon)\{ \textbf{0} \} | (mallet,\varphi_m,\epsilon)\{ \textbf{0} \} \\
Workers ::=  & !x(s).req(s)\{ hard\_hit(s) \} | !y(t).req(t)\{ soft\_hit(s) \} \\
Jobs ::=  & \bar{x}\langle hammer \rangle.\bar{y}\langle mallet \rangle.\bar{x}\langle mallet \rangle.\bar{x}\langle hammer \rangle.\textbf{0} \\
Workshop ::= & Tools | Workers | Jobs
\end{array}
\]
}
\end{example}


\paragraph{Operational semantics}
The operational semantics of our calculus is
defined by the transition relation given in 
Tab.~\ref{tab:sos}.
Labels $\mu,\mu'$ for transitions are $\tau$ for silent actions, $x(w)$ for free input, $\bar{x}v$ for free output,
$\bar{x}(v)$ for bound output, $\alpha(r)$, $ \alpha ? \textbf{r} $ and $ \overline{\alpha(r)} $ ($rel(r)$, $rel?r$ and $\overline{rel(r)}$, resp.) for closed, open and faulty access or release actions over resource $ r $.
The effect of bound output is to extrude the sent name from the
initial scope to the external environment.

\newcommand{\cano}[1]{\lfloor#1\rfloor}

We assume a notion of structural congruence and we denote it by $\equiv$.
This includes the standard laws of the $ \pi $-calculus, such as 
the monoidal laws for the parallel composition and 
the choice operator.
To simplify the definition of our Control Flow Analysis,
we impose a discipline in the choice of fresh names, and therefore to alpha-conversion.
Indeed, the result of analysing a process $P$, must still hold
for all its derivative processes $Q$, including all the processes obtained from $Q$ by alpha-conversion.
In particular, the CFA uses the names and the variables occurring in $P$.
If they were changed by the dynamic evolution, the analysis values would become a sort of dangling references, no more connected
with the actual values.
To statically maintain the identity
of values and variables, we partition all the
names used by a process into finitely many equivalence classes.
We denote with $\cano{n}$ the equivalence class of the name $n$, that is called \textit{canonical name} of $n$.
Not to further overload our
notation, we simply write $n$ for $\cano{n}$, when unambiguous.
We further demand that two names can be alpha-renamed only when they have
the same canonical name. 

In addition, we introduce specific laws for managing the 
resource-aware constructs, reported in Fig.~\ref{fig:cong}.
If two processes $P_1$ and $P_2$ are equivalent, then
also $P_1$ and $P_2$ when plugged inside the same resource boundaries are.
Resource request and resource joint points can be swapped with the 
restriction boundary since restriction is not applied to resource 
names but only to channel names. 
The last law is crucial for managing the discharge of resources.
This law allows rearrangements of available resources, e.g.~an available resource is allowed to enter or escape within a 
resource boundary. 

\begin{figure}
{\small
\[
\begin{array}{l}
  (\nu{}x)(r,\varphi,\eta) \lbrace P \rbrace  \equiv (r,\varphi,\eta) \lbrace (\nu{}x)P \rbrace \\
  (\nu{}x)req(r) \lbrace P \rbrace  \equiv req(r) \lbrace (\nu{}x)P \rbrace \\
  (r_2,{\varphi}_2,{\eta}_2) \lbrace \textbf{0} \rbrace \parallel (r_1,{\varphi}_1,\eta_1) \lbrace P \rbrace \equiv (r_1,{\varphi}_1,{\eta}_1) \lbrace (r_2,{\varphi}_2,{\eta}_2) \lbrace \textbf{0} \rbrace \parallel P \rbrace \\
  
\end{array}
\]
}
\caption{Structural congruence.} 
\label{fig:cong}
\end{figure}

%

\begin{table}[th]
{\small
\[
\begin{array}{llll}


{\rm (Act)} &
  { \pi.P \xrightarrow{\pi} P}\ \ \ \ \pi \neq \alpha(r),rel(r) &

{\rm (Cong)} &
  \irule{ P_1 \equiv P_1'\ \ P_1' \xrightarrow{\mu} P_2'\ \ P_2'\equiv P_2 }
  { P_1 \xrightarrow{\mu} P_2 }  \\

{\rm (Par)} &
  \irule{ P_1 \xrightarrow{\mu} P_1' }
  { P_1 \parallel P_2 \xrightarrow{\mu} P_1' \parallel P_2 }   \ \ 
\bn(\mu) \cap  \fn(P_2)= \emptyset\ \ \ &

{\rm (Choice)} &
  \irule{ P_1 \xrightarrow{\mu} P_1' }
  { P_1 + P_2 \xrightarrow{\mu} P_1'} \\

{\rm (Res)} &
  \irule{ P \xrightarrow{\mu} P'}
  { (\nu{}z)P  \xrightarrow{\mu} (\nu{}z)P' } \  z \not\in \n(\mu) &

{\rm (Open)} &
  \irule{ P \xrightarrow{\bar{x}y} P'}
  { (\nu{}y)P  \xrightarrow{\bar{x}(y)} P' }  \ y \neq x
  \\

{\rm (Comm)}&
  \irule{ P_1 \xrightarrow{\bar{x}y} P_1'\ \ P_2 \xrightarrow{x(z)} P_2' }
  {P_1 \parallel\ P_2 \xrightarrow{\tau} P_1' \parallel P_2' \lbrace 
y/z \rbrace } )&

{\rm (Close)} &
  \irule{ P_1 \xrightarrow{x(z)} P_1'\ \ P_2 \xrightarrow{\bar{x}(y)} P_2' }
  {P_1 \parallel\ P_2 \xrightarrow{\tau} (\nu{}y)(P_1' \parallel P_2'\lbrace 
y/z \rbrace)} \\

\\
\\
{\rm (Act_R)} &
\begin{array}{l}
\alpha(r).P \xrightarrow{\alpha?r} P \\
rel(r).P \xrightarrow{rel?r} P
\end{array}
&
{\rm (Comm_R)}&
  \irule{ P_1 \xrightarrow{\bar{x}r} P_1'\ \ P_2 \xrightarrow{x(s)} P_2' }
  {P_1 \parallel\ P_2 \xrightarrow{\tau} P_1' \parallel P_2' \lbrace 
r/s \rbrace } \\

  \\

\end{array}
\]

\[
\begin{array}{ll}
{\rm (Acquire)} &
  { req(r)\lbrace P \rbrace \parallel (r,\varphi,\eta)\lbrace \textbf{0} \rbrace \xrightarrow{\tau} (r,\varphi,\eta) \lbrace P \rbrace}  

\\
\\

{\rm (Release)} &
  \irule{P \xrightarrow{rel?r} P'}  { (r,\varphi,\eta)\lbrace P \rbrace \xrightarrow{rel(r)} (r,\varphi,\eta.rel) \lbrace \textbf{0} \rbrace \parallel P'}
\end{array}
\]

\[
\begin{array}{llll}

{\rm (Policy_1)} &
  \irule{P \xrightarrow{\alpha ? r} P'\ \ \ \eta.\alpha \models \varphi}
  { (r,\varphi,\eta) \lbrace P \rbrace \xrightarrow{\alpha(r)} 
(r,\varphi,\eta.\alpha) \lbrace P' \rbrace } 
  
&
  
{\rm (Policy_2)} &
  \irule{P \xrightarrow{\alpha ? r} P'\ \ \ \eta.\alpha \not\models \varphi}
  { (r,\varphi,\eta) \lbrace P \rbrace \xrightarrow{\overline{\alpha(r)}} 
(r,\varphi,\eta) \lbrace \textbf{0} \rbrace \parallel P' }\\

{\rm (Local_{1})} &
  \irule{P \xrightarrow{\mu} P'}  { (r,\varphi,\eta) \lbrace P \rbrace \xrightarrow{\mu} (r,\varphi,\eta) \lbrace  P' \rbrace }\ r \not \in \n(\mu) 
  
  &

{\rm (Local_{2})} &
  \irule{P \xrightarrow{\mu} P'\ } { req(r)\lbrace P \rbrace \xrightarrow{\mu} req(r)\lbrace P' \rbrace 
} \ r \not \in \n(\mu) 

\\
\\

{\rm (Appear)} &
{ P \xrightarrow{\tau} P \parallel (r,\varphi,\eta)\lbrace \textbf{0} \rbrace} 

&

{\rm (Disappear)} &
  { (r,\varphi,\eta)\lbrace P \rbrace \xrightarrow{\tau} \textbf{0}} \\

\end{array}
\]
}
\caption{Operational Semantics.} 
\label{tab:sos}
\end{table}

%
%
The rules $ Act $, $Par$, $Res$, $ Comm $, $Cong $, $Choice$, $ Open $ and $ Close $ are the standard $\pi$-calculus ones.
The rule $ Act $ describes actions of processes, e.g.~ the silent action, free input and free output. 
Concretely, $ \bar{x}w.P$ sends the name $w$ along the channel $x$ and then behaves like $P$, while $ x(w).P $ receives a name via the channel $x$, to which $w$ is bound, and then behaves like $P$. 
We only observe that our semantics is a late one, 
e.g.~$w$ is actually bound to a value when a communication occurs. 
Finally, $\tau.P$ performs the silent action $\tau$ and then behaves like $P$.

The rule $Par$ expresses the parallel computation of processes, while the rule $Choice$ represents a choice among alternatives.
The rule $Comm$ is used to communicate free names.
The rules $Res$ and $Open$ are rules for restriction. 
The first ensures that an action of $P$ is also an action of $(\nu{}z)P $, provided that the restricted name $z$ is not in the action.
In the case of $ z $ in the action, the rule $Open$ transforms a free output action $\bar{x}z$ into a bound output action $\bar{x}(z)$, which basically expresses opening scope of a bound name.
The rule $Close$ describes communication of bound names, which also closes the scope of a bound name in communication. 

We are now ready to comment on the semantic rules 
corresponding to the treatment of resources. 
The rule $ Act_R $ models a process 
that tries to perform an action $\alpha$ ($ rel $, resp.) on the resource $r$. This attempt
is seen as an {\em open action}, denoted by the label $ \alpha ? r $ ($ rel?r $, resp.). 

Intuitively, if the process is inside the scope of $r$ (see the rule $Local_1$), and the action satisfies the policy for $r$,
then the attempt will be successful and 
the corresponding action will be denoted by the label $\alpha(r)$ (see the rule $ Policy_1$).
If this is not the case, the process is stuck. 
Similarly, if the process tries to release a resource with the action $ rel $.
 

We introduce the rule $ Comm_R $ to model the communication of resource names between processes.

When a resource $r$ is available, then it 
can be acquired by a process $P$ that enters the corresponding
resource boundary $(r,\varphi,\eta)$, as stated by the rule $Acquire$.

Symmetrically, according to the rule $ Release $, the process $ P $ can 
release an acquired resource $r$ and update the state of its resources by appending $rel$ to $\eta$.
In the resulting process, the process $P$ escapes the resource boundary.
Furthermore, the resource becomes available, i.e.~it encloses the 
empty process $\textbf{0}$.
If the process is not inside the scope of $r$ (see the rule $Local_1$),
then, as in the case of accesses, the process is stuck. 

The rules $ Policy_1, Policy_2 $ check 
whether the execution of the action $\alpha$ on the 
resource $r$ obeys the policy $\varphi$, i.e.~whether the updated 
state $\eta.\alpha$, obtained by appending $\alpha $ to the current 
state $ \eta $, is consistent w.r.t.~$\varphi$.
If the policy is obeyed, then the updated state $\eta.\alpha$ is 
stored in the resource state according to the rule $ Policy_1 $ and the action becomes \textit{closed} 
and if not, then the resource is forcibly released according to the rule $ Policy_2 $ and the action becomes \textit{faulty}. 
Notice that $ Policy_2 $ is the rule managing the recovery from bad access to resources.

The rules $ Local_1 $ and $ Local_2 $ express that actions can bypass 
resource boundaries for $r$ only if they do not involve the resource 
$r$.

Finally, the rules $ Appear $ and $ Disappear $ describe the abstract behaviour of the resource manager performing asynchronous resource reconfigurations. 
In other words, resource configuration is not under the control of processes. 
Resources are created and destroyed by external entities and processes can only observe their presence/absence. 
This is formally represented by the rules $ Appear $ and $ Disappear $.

\begin{example}
To explain the operational semantics, we come back to our running example. The following trace illustrates how the workshop works. 
At the beginning, $Workers$ instantiates a new worker (a resource request point) when receiving a hard job: 
{
\[
\begin{array}{rl}
  & Workshop \\
  & \equiv  Workers | Tools | x(s).req(s)\{ hard\_hit(s) \} | \bar{x}\langle hammer \rangle.Jobs'  \\
  & \inter{\tau} Workers | Tools | Jobs' | req(hammer)\{ hard\_hit(hammer) \}, \\
\end{array}
\]
} 
where $ Jobs' ::= \bar{y}\langle mallet \rangle.\bar{x}\langle mallet \rangle.\bar{x}\langle hammer \rangle$. 
At this point the new worker can take a hammer and other jobs are also available (on the channel $x,y$). 
In the following, for the sake of simplicity, we only show sub-processes that involve computation. 
Assume that the new worker takes a hammer, then we have the following transition:
{
\[
\begin{array}{rl}
   & req(hammer)\{ hard\_hit(hammer) \} | (hammer,\varphi_h,\epsilon)\{ \textbf{0} \} \\
   & \inter{\tau}  (hammer,\varphi_h,\epsilon)\{ hard\_hit(hammer) \} \\
 \end{array}
\]
}
Now, three workers are similarly instantiated for doing all remaining jobs.
{
\[
\begin{array}{rl}
 & Workers |  Jobs' \\
 &  \inter{\tau} Workers |  req(mallet)\{ soft\_hit(mallet) \} | \bar{x}\langle mallet \rangle.\bar{x}\langle hammer \rangle \\
 &  \inter{\tau} Workers |  req(mallet)\{ soft\_hit(mallet) \} | req(mallet)\{ hard\_hit(mallet). \} | \bar{x}\langle hammer \rangle \\
 &  \inter{\tau} Workers |  req(mallet)\{ soft\_hit(mallet) \} | req(mallet)\{ hard\_hit(mallet) \} | req(hammer)\{ hard\_hit(mallet) \} \\
\end{array}
\]
}

In the current setting, the new three workers make one request on the remaining hammer and two requests on the mallet. 
Since we have only one mallet, one of two mallet requests could be done at a time. 
Suppose the first job get done first, we have the following transition:
{
\[
\begin{array}{rl}
   &  (hammer,\varphi_h,\epsilon)\{ hard\_hit(hammer) \} \\
   & \xrightarrow{hard\_hit(hammer)}  (hammer,\varphi_h,hard\_hit)\{ \textbf{0} \}

\end{array}
\]
}
Note that the hammer is available again. 
Similarly, the second job is done as follows:
{
\[
\begin{array}{rl}
  & req(mallet)\{ soft\_hit(mallet) \} | (mallet,\varphi_m,\epsilon)\{ \textbf{0} \} \\
  & \inter{\tau} (mallet,\varphi_m,\epsilon)\{ soft\_hit(mallet) \} \\   
  & \xrightarrow{soft\_hit(mallet)} (mallet,\varphi_m,soft\_hit)\{ \textbf{0} \}
\end{array}
\]
}
If the third job would be processed, then a forced release could occur. 
This happens because the worker attempts to do a hard hit by using a mallet in doing the job, which violates the mallet policy.
{
\[
\begin{array}{rl}
 & req(mallet)\{ hard\_hit(mallet).\textbf{0} \} | (mallet,\varphi_m,\epsilon)\{ \textbf{0} \} \\
 & \inter{\tau} (mallet,\varphi_m,\epsilon)\{ hard\_hit(mallet) \} \\
 & \xrightarrow{\overline{hard\_hit(hammer)}} (hammer,\varphi_h,\epsilon)\{ \textbf{0} \} | \textbf{0}
\end{array}
\]
}
Finally, the similar trace is for the fourth job.
\end{example}

\section{Control Flow Analysis}

\newcommand{\FORM}[2]
{\mbox{$(\rho,\kappa,\Gamma,\Psi) \models^{#1}{#2}$}}
\newcommand{\FORMN}[2]
{\mbox{$(\rho,\kappa,\Gamma,\Psi) \not\models^{#1}{#2}$}}
\newcommand{\FORMPRIM}[2]
{\mbox{$(\rho',\kappa',\Gamma',\Psi') \models^{#1}{#2}$}}
\newcommand{\FORMJ}[2]
{\mbox{$(\rho_j,\kappa_j,\Gamma_j,\Psi_j) \models^{#1}{#2}$}}
\newcommand{\EST}
{\mbox{$(\rho,\kappa,\Gamma,\Psi)$}}
\newcommand{\ESTPRIM}
{\mbox{$(\rho',\kappa',\Gamma',\Psi')$}}
\newcommand{\ESTJ}
{\mbox{$(\rho_j,\kappa_j,\Gamma_j,\Psi_j)$}}
\newcommand{\IN}{\hbox{\;\scriptsize{\sf E}\;}}

In this section, we present a CFA for our calculus, extending the one for $ \pi $-calculus~\cite{DBLP:journals/iandc/BodeiDNN01}. 
The CFA computes a safe
over-approximation of all the possible communications of resource and channel names on channels.
Furthermore, it provides an over-approximation of all the possible usage traces on the given resources and records the names of the resources
that can be possibly not released, thus providing information on possible bad usages.
The analysis is performed under the perspective of processes. 
This amounts to saying that the analysis tries to answer the following question: 
``Are the resources initially granted sufficient to guarantee a correct usage?''.
In other words, we assume that a certain fixed amounts of resources is given and 
we do not consider any dynamic reconfiguration, possible in our calculus, 
due to the rules $ Appear $ and $ Disappear $.
The reconfiguration is up to the resource manager and is not addressed by the CFA.

For the sake of simplicity, we provide the analysis for a subset of our calculus, in which processes enclosed in the scopes of resources are {\em sequential processes} (ranged over by $ Q,Q' $), as described by the following syntax. 
Intuitively, a sequential process represents a single thread of execution in which one or more resources can be used.


{\small
\[
\begin{array}{lrlllrll}
P,P' &::=  & \hbox{as before in Fig.\ref{fig:syn}} & 
        \ \
        & Q,Q' &::=\ & sequential\ processes & \\

& \;\mid\; & (r,\varphi,\eta) \lbrace Q \rbrace &
& & & \textbf{0} &  \\

& \;\mid\; & req(s) \lbrace Q \rbrace & 
& & \;\mid\; & (\nu{}z)\ Q & \\

&  &  & 
& & \;\mid\; & \pi.Q & \\

&  &  & 
& & \;\mid\; & Q + Q' & \\

&  &  & 
& & \;\mid\; & (r,\varphi,\eta) \lbrace Q \rbrace & \\

&  &  & 
& & \;\mid\; & (r,\varphi,\eta) \lbrace \textbf{0} \rbrace || Q & \\

&  &  &
& & \;\mid\; & req(s) \lbrace Q \rbrace & \\

%
%
%
%

\end{array}
\]
}

This implies that one single point for releasing each resource occurs in each non deterministic branch of a process. 
The extension to general parallel processes is immediate. Nevertheless, it requires some more complex technical machinery 
in order to check whether all the parallel branches synchronise among them, 
before releasing the shared resource.

In order to facilitate our analysis, 
we further associate labels $ \chi \in \mathcal{L} $ 
with resource boundaries as follows: 
$ (r,\varphi,\eta) \{ Q \}^{\chi} $ and $ req(r) \{ Q \}^{\chi} $,
in order to give a name to the sub-processes in the resource scopes.
Note that this annotation can be performed in a pre-processing step and does not affect the semantics of the calculus. 
During the computation, resources are released and acquired by other processes.
Statically, sequences of labels $S \in \mathcal{L}^* $ are used to record the sequences of sub-processes possibly entering the scope
of a resource.
Furthermore, to 
make our analysis more informative, we enrich the execution traces $\eta$ 
with special actions that record the fact that a resource has been possibly:
\begin{itemize}
\item acquired by the process labelled $\chi$: $in(\chi)$, with a successful request;
\item released by the process labelled $\chi$: $out(\chi)$ with a successful release;
\item taken away from the process labelled $\chi$: $err\_out(\chi)$ 
because of an access action on $r$ that does not satisfy the policy. 
\end{itemize}
The new set of traces is $ \hat{\mathcal{A}}^* $, where $ \hat{\mathcal{A}} = \mathcal{A} \cup \{in(\chi),out(\chi),err\_out(\chi) \ | \ \chi \in \mathcal{L}\} $. 
The corresponding dynamic traces can be obtained by simply removing all the special actions.

The result of analysing a process $P$ is a tuple $ (\rho,\kappa,\Gamma,\Psi)$ called {\em estimate} of $P$, that provides an
approximation of resource behavior.
More precisely, 
$\rho$ and $\kappa$ offer an over-approximation of all the possible values that the variables in the system may
be bound to, and of the values that may flow on channels.
The component 
$ \Gamma $ provides a set of traces of actions on each resource. Finally, $ \Psi $ records a set of the resources that can be possibly not released. 
Using this information, we can statically check resource usages against the required policies. 

To validate the correctness of a given estimate $ (\rho,\kappa,\Gamma,\Psi)$, we state a set of clauses that operate upon judgments 
in the form $\FORM{\delta}{P}$, where $ \delta $ is a sequence of pairs $ [(r,\varphi,\eta),S] $,
recording the resource scope nesting. 
This sequence is initially empty, denoted by $ [\epsilon,\epsilon] $.

The analysis correctly captures the behavior of $P$, i.e.~the estimate $ (\rho,\kappa,\Gamma,\Psi)$ is valid for all the derivatives $P'$ of $P$. 
In particular, the analysis keeps track of the following information:
\begin{itemize}
\item An approximation $ \rho: \mathcal{N} \cup \mathcal{R} \rightarrow  \wp(\mathcal{N} \cup \mathcal{R})$ of names bindings. 
If $a \in \rho(x)$ then the channel variable $x$ can assume the channel value $a$. 
Similarly, if $r \in \rho(s)$ then the resource variable $s$ can assume the resource value $r$.

\item An approximation $ \kappa: \mathcal{N} \rightarrow \wp(\mathcal{N} \cup \mathcal{R}) $ of the values 
that can be sent on each channel. If $b \in \kappa(a)$,
then the channel value $b$ can be output on the channel $a$, while $r \in \kappa(a)$, then the resource value $r$ can be output on the channel $a$.


\item An approximation $ \Gamma: \mathcal{R} \rightarrow \wp(\{ [(\varphi,\eta),S] | \ \varphi \in \Phi, S \in \mathcal{L}^*,
\eta \in \hat{\mathcal{A}}^* \} )$ of resource behavior.
If $[(\varphi,\eta),S] \in \Gamma(r)$ then $\eta$ is one of the possible traces over $ r $ that is performed by a sequence of sub-processes, whose labels $\chi$ are
juxtaposed in $ S $.


\item An approximation $ \Psi \in \wp(\{ \delta \ | \  \delta \mbox{ is a sequence of pairs }  [(r,\varphi,\eta),S] \}$
of the resources which are possible locked by processes in deadlock for trying to access or to release a resource not in their scope. 
More precisely, if $\delta$ is in $\Psi$ and $[(r, \phi,\eta), S]$ occurs in $\delta$, then the resource $r$ can be possibly acquired 
by a process that can be stuck and that therefore could not be able to release it.

\end{itemize}

The judgments of the CFA are given in Tab.~\ref{cfa}, which are based on structural induction of processes. 
We use the following
shorthands to simplify the treatment of the sequences $\delta$.
The predicate $[(r,\varphi,\eta),\chi] \IN \delta$ is used to check whether the pair $[(r,\varphi,\eta),\chi]$ occurs in $\delta$, 
i.e.~whether $\delta = \delta' [r,(\varphi,\eta),\chi] \delta''$. 
With $ \delta\{[(r,\varphi,\eta.\alpha),S]/[(r,\varphi,\eta),S]\} $ we indicate that the pair $ [(r,\varphi,\eta),S] $ is replaced by $ [(r,\varphi,\eta.\alpha),S] $ in the sequence $\delta$. 
With $\delta \setminus [(r,\varphi,\eta),S]$ we indicate the sequence 
where the occurrence $[(r,\varphi,\eta),S]$ has been removed, i.e.~the sequence $\delta'\delta''$,
if $\delta = \delta' [(r,\phi,\eta),S] \delta''$.

\begin{table}
\begin{tabular}{| l  l |}
\hline
 
 & \\
  
\FORM{\delta}{\textbf{0}}
 & 
{\rm\ iff\ true} \\

 & \\

\FORM{\delta}{\tau.P} 
 & 
${\rm\ iff\ } \FORM{\delta}{P}$\\

 & \\

\FORM{\delta}{\bar{x}w.P} 
 & ${\rm\ iff\ } \forall a \in \rho(x): \rho(w) \subseteq \kappa(a) \ \wedge \ \FORM{\delta}{P}$ \\
 
 & \\
 
\FORM{\delta}{x(y).P} 
 & $ {\rm\ iff\ } \forall a \in \rho(x): \kappa(a) \cap \mathcal{N} \subseteq \rho(y) \ \wedge \ \FORM{\delta}{P} $\\

 & \\
 
\FORM{\delta}{x(s).P} 
 & 
${\rm\ iff\ } \forall a \in \rho(x): \kappa(a) \cap \mathcal{R} \subseteq \rho(s) $ \\ 
& $ \ \wedge \  \forall r \in \rho(s): \FORM{\delta}{P\{r/s\}} $ \\

 & \\ 
 
\FORM{\delta}{P_1 + P_2}
 & 
${\rm\ iff\ } \FORM{\delta}{P_1} \wedge \FORM{\delta}{P_2}$ \\ 

 & \\
 
\FORM{\delta}{P_1 \parallel P_2}
 & 
${\rm\ iff\ } \FORM{\delta}{P_1} \wedge \FORM{\delta}{P_2}$ \\ 

 & \\
 
\FORM{\delta}{(\nu{}x)P}
 & 
${\rm\ iff\ } \FORM{\delta}{P} \wedge x \in \rho(x)$ \\
 
 & \\
 
\FORM{\delta}{!P} 
 & 
{\rm\ iff\ } \FORM{\delta}{P} \\ 

 & \\

\FORM{\delta}{(r,\varphi,\eta) \{ Q \}^{S}}
 & 
{\rm\ iff\ } \FORM{\delta[(r,\varphi,\eta), S ]}{Q} \\
 
 & \\
 
 \FORM{\delta}{(r,\varphi,\eta) \{ {\bf 0} \}^{S}}
 & 
${\rm\ iff\ } [(\varphi,\eta),S] \in \Gamma(r)$ \\

 & \\
 

\FORM{\delta}{req(r) \{Q \}^{\chi}}
 & 
${\rm \ iff\ } \forall [(\varphi,\eta),S] \in \Gamma(r) \wedge \chi \not \in S
\wedge $
\\
 &\ \ \  $\Rightarrow \FORM{\delta[(r,\varphi,\eta.in(\chi)),S\chi]}{Q}$\ \\

 & \\
 
  
\FORM{\delta}{\alpha(r).Q} 
 & 
 ${\rm\ iff\ }\ [(r,\varphi,\eta),S\chi] \IN \delta \wedge \eta.\alpha \models \varphi \Rightarrow \FORM{\delta'}{Q}\ $ \\
 &\ \ \ $\wedge\ [(r,\varphi,\eta),S\chi] \IN \delta \wedge \eta.\alpha \not \models \varphi $\\
 &\ \ \ \ \ \ \ $\Rightarrow [(\varphi,\eta.err\_out(\chi)),S\chi] \in \Gamma(r) \wedge \FORM{\delta''}{Q} $ \\
 &\ \ \ $\wedge\ [(r,\varphi,\eta),S\chi] \not\IN \delta  \Rightarrow  \delta \in \Psi$ \\
 &  $\mbox{ with } \delta' = \delta\{ [(r,\varphi,\eta.\alpha),S\chi]/[(r,\varphi,\eta),S\chi] \}$  \\
 &  $\mbox{ and } \delta'' = \delta \setminus [(r,\varphi,\eta),S\chi]$  \\
 & \\

\FORM{\delta}{\omega(r).Q} 
 & 
 ${\rm\ iff\ }\ [(r,\varphi,\eta),S\chi] \IN \delta \Rightarrow \FORM{\delta \setminus [(r,\varphi,\eta),S\chi]}{Q}\ $\\
 &\ \ \ $\wedge\ [(\varphi,\eta.\omega.out(\chi)),S\chi] \in \Gamma(r)$ \\
 &\ \ \ $\wedge\ [(r,\varphi,\eta),S\chi] \not\IN \delta \Rightarrow  \delta \in \Psi $ \\

 & \\

\FORM{\delta}{(r,\varphi,\eta) \{ {\bf 0} \}^{S} \parallel Q}
 & 
{\rm \ iff\ } 
\FORM{\delta}{(r,\varphi,\eta) \{ {\bf 0} \}^{S}} $\wedge $ \FORM{\delta}{Q}\ \\

  & \\
  
\hline

\end{tabular}
\caption{CFA Equational Laws}
\label{cfa}
\end{table}

All the clauses dealing with a compound process check that the analysis also
holds for its immediate sub-processes. In particular, the analysis of $!P$ and that
of $(\nu x)P$ are equal to the one of $P$. This is an obvious source of imprecision
(in the sense of over-approximation).
We comment on the main rules.
Besides the validation of the continuation process $P$, the rule for output,
requires that the set of names that can be communicated along
each element of $\rho(x)$ includes the names to
which $y$ can evaluate.
Symmetrically, the rules for input demands that the set of 
names that can pass along $x$ is included in the set of names to
which $y$ can evaluate.
Intuitively, the estimate components take into account the possible dynamics
of the process under consideration. The clauses' checks mimic the semantic evolution,
by modelling the semantic preconditions and the consequences of the possible
synchronisations. In the rule for input, e.g., CFA
checks whether the precondition of a synchronisation is 
satisfied, i.e.~whether there is a corresponding output possibly sending a value that can be received
by the analysed input. 
The conclusion imposes the additional requirements
on the estimate components, necessary to give a valid prediction of the analysed
synchronisation action, mainly that the variable $y$ can be bound to that value.

To gain greater precision in the prediction of resource usages, in the second rule,
the continuation process is analysed, for all possible bindings of the resource variable $s$.
This explains why we have all the other rules for resources, without resource variables.

The rule for \textit{resource joint point} updates $ \delta $ to record that 
the immediate sub-process is inside the scope of the new resource and there it is analysed. 
If the process is empty, i.e.~in the case the resource is available, the trace of actions is recorded in $ \Gamma(r)$.

In the rule for \textit{resource request point}, the analysis for $Q$
is performed for every possible element $ [(\varphi,\eta),S] $ from the component $ \Gamma(r)$. 
This amounts to saying that the resource $r$ can be used starting from any possible previous trace $\eta$.
In order not to append the same trace more than once, we have the condition that $ S $ does not contain $ \chi$.
This prevents the process labelled $\chi$ to do it.
Furthermore, $\eta$ is enriched by the special action $in(\chi)$ that records the fact that the resource $r$ can be possibly acquired by the process labelled $\chi$.


According to the rule for \textit{access action}, if the pair $ [(r,\varphi,\eta),S\chi] $ occurs in $ \delta $ (i.e.~if we are inside the resource scope of $r$) and the updated history $ \eta.\alpha $ obeys the policy $ \varphi $, then the analysis result also holds for the immediate subprocess and $ \delta $ is updated in $\delta'$, by replacing
$ [(r,\varphi,\eta),S\chi] $ in $\delta$ with $ [(r,\varphi,\eta.\alpha),S\chi]$, 
therefore recording the resource accesses to $r$ possibly made by the sub-process labelled by $\chi$.

In case the action possibly violates the policy associated with $r$ (see the last conjunct), the process labelled $\chi$ may loose the resource $r$, as recorded by the trace in $\Gamma$, 
$[(\varphi,\eta.err\_out(\chi)),S\chi]$, with the special action $err\_out(\chi)$ appended to $\eta$.
If instead, the action on $r$ is not viable because the process is not in the scope of $r$, then all the resources in the context $\delta$ could not be released, as recorded by the component $\Psi$.

According to the rule for {\em release}, the trace of actions $\eta' = \eta.\omega.out(\chi)$ over $ r $ at $ \chi $ is recorded in $ \Gamma(r) $. 
Other sub-processes can access the resource starting from the trace $\eta'$. 
Furthermore, $ [(r,\varphi,\eta),S] $ is removed from $ \delta$ and this reflects the fact that the process $Q$ can exit its scope, once released the resource $r$.
Similarly, in the last rule, $ [(r,\varphi,\eta),S] $ is removed from $ \delta$ and there
the process $Q$ is analysed.
Again, if the action on $r$ is not possible because the process is not in the scope of $r$, then all the resource in the context $\delta$ could not be released, as recorded by the component $\Psi$.

\begin{example}
We briefly interpret the results of CFA on our running example. A more complex of exemplification of CFA is given in the next example  (see below). First we associate labels with the resource boundaries as follows:
{
\[
\begin{array}{rl}
Tools ::= & (hammer,\varphi_h,\epsilon)\{ \textbf{0} \}^{\chi_1} | (hammer,\varphi_h,\epsilon)\{ \textbf{0} \}^{\chi_2} | (mallet,\varphi_m,\epsilon)\{ \textbf{0} \}^{\chi_3} \\
Workers ::=  & !x(s).req(s)\{ hard\_hit(s) \}^{\chi_h} | !y(t).req(t)\{ soft\_hit(s) \}^{\chi_m} \\
\end{array}
\]
}
It is easy to see that there is one policy violation, which is captured by our CFA in the component $\Gamma(hammer)$, from which we can extract the following trace: $(in(\chi_m).err\_out(\chi_m),\chi_m)$. 
It occurs when doing the third job the worker tries to hit \textit{hard} using a mallet. 
We know that the channel $x$ ($y$, resp.) is supposed to send/receiving hard jobs (soft jobs, resp.), i.e.~sending/receiving $hammer$ ($mallet$, resp.) and names $s$ and $t$ are supposed to be bound to $hammer$ and $mallet$ respectively. 
By checking the component $\rho$ and $\kappa$, we can explain the above violation too.
On the one hand, we found that $\rho(t)$ is a singleton set of $mallet$, while $\rho(s)$ is a set of $hammer$ and $mallet$, which is a wrong bound of $s$.
On the other hand, similarly we found that $\kappa(x)$ contains only $hammer$, while $\kappa(y)$ contains $hammer$ and $mallet$, which is a wrong use of $y$. 
\end{example}

\begin{example}[Robot Scenario]
We now consider a 
scenario, where a set of robots collaborate to reach a certain goal, 
e.g.~to move an item from one position to another.
Without loss of generality, we assume that robots operate in a space 
represented by a two-dimensional grid.
We also assume that certain positions over the grid are faulty, and 
therefore they cannot be crossed by robots.
To move the item, a robot needs to take it, and this is allowed 
provided that the
item is co-located within the range of robot's sensor.
Moreover, since robots have a small amount of energy power, they can 
perform just a few of steps with the item.
Finally, we consider three families of robots ($R_1, R_2$ and $R_3$): 
each robot in the family has different computational 
capabilities.

\begin{figure}
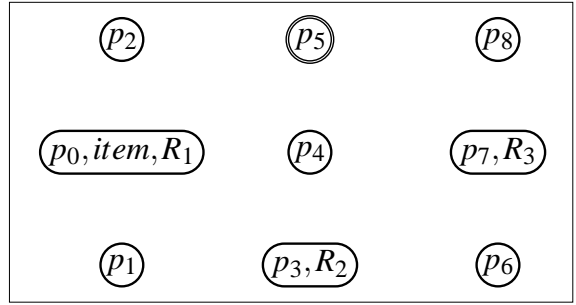


\centering
\SmallPicture\VCDraw{%
\ShowFrame
\begin{VCPicture}{(-2,0)(13,8)}
\LargeState
\Large{}
\State[p_1]{(1,1)}{A1}
\StateVar[p_0,item,R_1]{(1,4)}{A2}
\State[p_2]{(1,7)}{A3}

\StateVar[p_3,R_2]{(6,1)}{B1}
\State[p_4]{(6,4)}{B2}
\FinalState[p_5]{(6,7)}{B3}

\State[p_6]{(11,1)}{C1}
\StateVar[p_7,R_3]{(11,4)}{C2}
\State[p_8]{(11,7)}{C3}



\end{VCPicture}


%
%
%
%
%
%
%
%
%
}

\caption{The initial configuration of the robot scenario.} \label{fig:robot}

\end{figure}
\begin{figure}
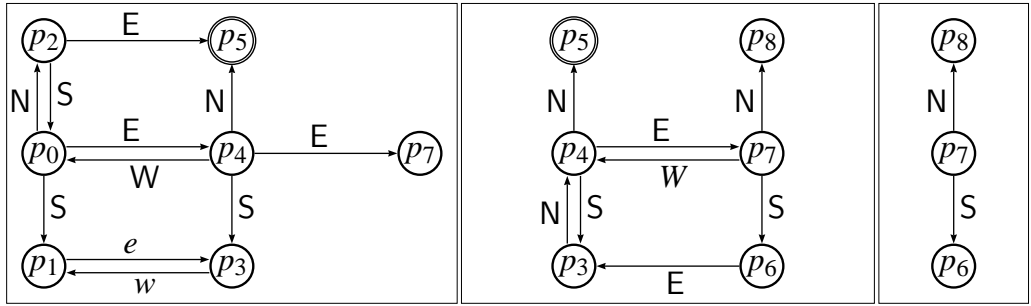


\SmallPicture\VCDraw{%
\ShowFrame


\begin{VCPicture}{(-2,0)(10,8)}
\LargeState
\Large{}
\State[p_1]{(-1,1)}{A1}
\State[p_0]{(-1,4)}{A2}
\State[p_2]{(-1,7)}{A3}

\State[p_3]{(4,1)}{B1}
\State[p_4]{(4,4)}{B2}
\FinalState[p_5]{(4,7)}{B3}

\State[p_7]{(9,4)}{C2}

\ForthBackOffset \EdgeL{A1}{B1}{e} \EdgeL{B1}{A1}{w}
\RstEdgeOffset

\ForthBackOffset
\EdgeL{A2}{A3}{\NR}
\EdgeL{A3}{A2}{\SD}
\RstEdgeOffset

\ForthBackOffset
\EdgeL{A2}{B2}{\E}
\EdgeL{B2}{A2}{\W}
\RstEdgeOffset

\EdgeL{B2}{C2}{\E}

\EdgeL{B2}{B1}{\SD}

\EdgeL{B2}{B3}{\NR}
\EdgeL{A3}{B3}{\E}
\EdgeL{A2}{A1}{\SD}
\end{VCPicture}

\begin{VCPicture}{(10,0)(21,8)}

\LargeState
\Large{}
\Large{}
\State[p_3]{(13,1)}{B1}
\State[p_4]{(13,4)}{B2}
\FinalState[p_5]{(13,7)}{B3}

\State[p_6]{(18,1)}{C1}
\State[p_7]{(18,4)}{C2}
\State[p_8]{(18,7)}{C3}




\ForthBackOffset
\EdgeL{B2}{C2}{\E}
\EdgeL{C2}{B2}{W}
\RstEdgeOffset

\ForthBackOffset
\EdgeL{B2}{B1}{\SD}
\EdgeL{B1}{B2}{\NR}
\RstEdgeOffset

\EdgeL{B2}{B3}{\NR}
\EdgeL{C2}{C3}{\NR}
\EdgeL{C1}{B1}{\E}
\EdgeL{C2}{C1}{\SD}
\end{VCPicture}

\begin{VCPicture}{(22,0)(26,8)}
\LargeState
\Large{}


\State[p_6]{(24,1)}{C1}
\State[p_7]{(24,4)}{C2}
\State[p_8]{(24,7)}{C3}






\EdgeL{C2}{C3}{\NR}
\EdgeL{C2}{C1}{\SD}
\end{VCPicture}

}

\caption{The policy automata of the robots' families: $ R_1 $ (left), 
$ R_2 $ (middle) and $ R_3 $ (right).} \label{fig:policies}

\end{figure}

Fig.~\ref{fig:robot} gives a pictorial description of the initial configuration of the scenario. 
Positions are represented by circles and double circles. Double 
circles indicate faulty positions.
The item is located at position $ p_0 $ and the goal is to move it
into the position $ p_8 $.
There is just one faulty position $ p_5 $, crossing through which is 
considered a failure.
Moreover, we consider a scenario where the three families of robots 
$R_1, R_2$ and $R_3$ are initially located at $ p_0 $, $ p_3 $ and 
$ p_7 $, respectively
(e.g. all the robots of the family $R_1$ are located at $ p_0 $).

Sensors are modelled by clearly identified resources. The sensor $j^{th}$ of 
the $i^{th}$ robot family is specified by the resource 
$ (sns_{i,j},\varphi_j,\eta_{i,j}) $, where
$ sns_{i,j} $ is the name of the sensor, $\eta_{i,j}$ is the abstract representation of the sequence of moving actions which led the robot from its 
initial position to the current one and initially equals to $ \epsilon $,
and $ \varphi_j$ is the global policy on demand.
We assume that each family of robots has its own policy described by 
the automata in Fig.~\ref{fig:policies}. The policy 
constraints robots' movement in the grid.
We model the movement activities of robots with the following 
actions: $ \E(sns) $, $ \W(sns) $, $ \SD(sns) $, and $ \NR(sns) $ 
that describe the movements on east (west, south and north, resp.).
Basically, sensors are a sort of private resources of the robots 
(each robot will never release its sensor) and the actions over 
sensors update their states.



The item is modelled by a resource of the form $ (IT,\varphi_{I},\eta) $, 
where $ \eta $ describes the sequence of actions performed on the item, 
and $ \varphi_{I} $ simply states that the item is never located at 
the position $ p_5 $. Initially, $ \eta $ is equal to $\epsilon$.
The same set of actions adopted for robots' movement (namely $ \E(IT) 
$, $ \W(IT) $, $ \SD(IT) $, and $ \NR(IT) $) are exploited to 
transport the item in the grid. 
Finally, each robot in the family $i \in \{1, 2, 3\}$ is specified by 
a process $ R_{i,j} $ of the form: $ (sns_{i,j},\varphi_j,\eta_{i,j}) \{ Q_{i,j} \}^{\chi} $, where $ Q_{i,j} $ specifies the  $j^{th}$ robot's behaviour of the  $i^{th}$ robot family and $\chi$ is a label associated with the resource boundary.
For instance, in the process $ Q_{2,3} $ (see below), the robot goes 
to north (without the item), then it tries to grasp the item. If this 
operation succeeds, the robot goes to east and releases the item there. 
Note that we use two monadic actions to move the item and the sensor together. 
This could be done by using polyadic actions, which however we leave for future work. 

For the sake of simplicity, we do not model co-location of sensors and items.
The specification of the robot scenario is given below.
{
\small
\[
\begin{array}{l}
  R_{1,1} :=  (sns_{1, 1},\varphi_1,p_0) \{ req(IT) \{ \E(IT).\E(sns_{1,1}).\SD(IT).\SD(sns_{1,1}).rel(IT)\}^{\chi_{r11}} \}^{\chi_{s11}} \\
  R_{1,2} :=  (sns_{1, 2},\varphi_1,p_0) \{ req(IT) \{ \E(IT).\E(sns_{1,2}). \E(IT).\E(sns_{1,2}).rel(IT) \}^{\chi_{r12}} \}^{\chi_{s12}} \\
  R_{1,3} :=  (sns_{1, 3},\varphi_1,p_0) \{req(IT) \{ \E(IT).\E(sns_{1,3}).rel(IT) \}^{\chi_{r13}} \}^{\chi_{s13}} \\
  R_{2,1} :=  (sns_{2, 1},\varphi_2,p_3) \{ req(IT) \{ \NR(IT).\NR(sns_{2,1}).\E(IT).\E(sns_{2,1}).rel(IT) \}^{\chi_{r21}} \}^{\chi_{s21}} \\
  R_{2,2} := (sns_{2, 2},\varphi_2,p_3) \{ req(IT) \{ \NR(IT).\NR(sns_{2,2}).\NR(IT).\NR( sns_{2,2}).rel(IT) \}^{\chi_{r22}} \}^{\chi_{s22}} \\
  R_{2,3} := (sns_{2, 3},\varphi_2,p_3) \{ NR(sns_{2,3}).req(IT) \{ \E(IT).\E( sns_{2,2}).rel(IT) \}^{\chi_{r23}} \}^{\chi_{s23}} \\
  R_{3,1} :=  (sns_{3, 1},\varphi_3,p_7) \{   req(IT) \{ \SD(IT).\SD(sns_{3, 1}).rel(IT) \}^{\chi_{r31}}  \}^{\chi_{s31}} \\
  R_{3,2} := (sns_{3, 2},\varphi_3,p_7) \{   req(IT) \{ \NR(IT).\NR(sns_{3, 2}).rel(IT) \}^{\chi_{r32}}  \}^{\chi_{s32}} \\
  \\
  System := (IT,\varphi_{I},p_0) \{\textbf{0} \}^{\chi_{IT}} \parallel R_{1,1} 
\parallel R_{1,2}  \parallel R_{1,3}  \parallel R_{2,1}  \parallel 
R_{2,2}  \parallel R_{2,3}  \parallel R_{3,1}  \parallel R_{3,2}
\end{array}
\]
}
The following trace illustrates the behaviour of the specification of the scenario. 
At the beginning, the item lies in the range of the family of robot $R_1$. 
Then a reconfiguration step putting together the robot $R_{1,1}$ and the item is performed.

{
\[
\begin{array}{l}
  System := (IT,\varphi_{I},\epsilon) \{\textbf{0} \} || (sns_{1,1},\varphi_1, \epsilon) \{ Q_{1,1} \} 
|| R_{1,2} || R_{1,3}  || R_{2,1}  || R_{2,2} || R_{2,3} || R_{3,1} || R_{3,2}
\equiv \\
\ \ \ \ \ \ \ \ \ \ \ \ \ \ (sns_{1,1},\varphi_1, \epsilon) \{ (IT,\varphi_I, \epsilon) \{ \textbf{0} \} 
|| Q_{1,1} \} || R_{1,2} || R_{1,3} || R_{2,1}  || R_{2,2} || R_{2,3} || R_{3,1}  || R_{3,2}
\end{array}
\]
}

As a result, robot $ R_{1,1} $ can grasp (acquire) the item; 
the pair item-robot moves on east, then on south. Finally, the robot disposes the item at the position $p_3$.
{
\[
\begin{array}{ll}
System \xrightarrow{\tau}
&
(sns_{1,1},\varphi_1, p_0) \{ (IT,\varphi_I, \epsilon) \{ Q_{1,1} \} || R_{1,2} || R_{2,1}  || R_{2,2} || R_{3,1} || R_{3,2} \\

& \xrightarrow{E(IT)}\xrightarrow{E(sns_{1,1})}
\xrightarrow{S(IT)}\xrightarrow{S(sns_{1,1})} 
\xrightarrow{rel(IT)} \\
& (IT,\varphi_I,\epsilon.E.S.rel) \{ \textbf{0} \} || (sns_{1,1},\varphi_1,\epsilon.E.S) \{ \textbf{0} \} || R_{1,2} || R_{2,1} || R_{2,2} || R_{3,1}  || R_{3,2} \\

\end{array}
\]
}

It is easy, given an initial location, to map a sequence of actions performed over the item into a path on the grid, namely each action operated over the item (i.e.~$ \E(IT) $, $ \W(IT) $, $ \SD(IT) $, and $ \NR(IT) $) corresponds to a single moving step in the space grid. The release action, instead, is interpreted as a sort of self-loop in the grid, i.e. the execution of the release action does not move the item.
For example, the sequence $ \epsilon.E.S.$ in the above setting would model the path $ p_0p_4p_3$. 
From now on, by abuse of notation, we will freely use paths in place of sequences of actions over the item/sensors.

Now, the item is in the range of the family of robots $ R_2 $. Again 
by applying the reconfiguration step, robot $R_{2,1}$ is allowed to 
operate with the item.
Then, it takes the item, makes a move on north, then on east, and 
disposes the item at the position $p_7$. For the sake of simplicity, in the following we show only sub-processes of the system that involve computation:
{
\[
\begin{array}{l}
(IT,\varphi_I,p_0 p_4 p_3 p_3) \{ \textbf{0} \} ||  R_{2,1} \\

\xrightarrow{\tau}
\xrightarrow{N(IT)}\xrightarrow{N(sns_{2,1})}
\xrightarrow{E(IT)}\xrightarrow{E(sns_{2,1})} 
\xrightarrow{rel(IT)} \\ 

(IT,\varphi_I,p_0 p_4 p_3 p_3 p_4 p_7 p_7) \{ \textbf{0} \} || 
(sns_{2,1},\varphi_2, p_3 p_4 p_7) \{ \textbf{0} \}  
\end{array}
\]
}
Note that a forced release would have occurred at this step if the item proceeded 
governed by the robot $ R_{2,2} $. 
The reason is that $ R_{2,2} $ attempts to move the item into the position $ p_5 $ 
and this results in releasing the item at the position $ p_4 $ by the rule $ Policy_2 $.
Now the robot $R_{3,2}$ has the chance to take the item, and,
if the north move occurs, the goal is achieved and the task is completed.

{
\[
\begin{array}{l}
(IT,\varphi_I, p_0 p_4 p_3 p_3 p_4 p_7 p_7) \{ \textbf{0} \} || R_{3,2} \\

\xrightarrow{\tau}
\xrightarrow{N(IT)}\xrightarrow{N(sns_{3,2})}
\xrightarrow{rel(IT)} \\ 

(IT,\varphi_I,p_0 p_4 p_3 p_3 p_4 p_7 p_7 p_8 p_8) \{ \textbf{0} \} || (sns_{3,2},\varphi_3, p_7 p_8)  \{ \textbf{0} \} \\

\end{array}
\]
}

%
%
Now we explain the features of the CFA. 
The CFA (in particular the $\Gamma$ component) computes the set of possible traces of the trajectories in the grid reaching the goal, among which the ones below:

{\small
\[
\begin{array}{l}
 in(\chi_{r11}).E.S.rel.out(\chi_{r11}).in(\chi_{r21}).N.E.rel.out(\chi_{r21}).in(\chi_{r32}).N.rel.out(\chi_{r32}),\chi_{r11}.\chi_{r21}.\chi_{r32} \\
 in(\chi_{r11}).E.E.rel.out(\chi_{r11}).in(\chi_{r32}).N.rel.out(\chi_{r32}),\chi_{r12}.\chi_{r32} \\
 in(\chi_{r13}).E.rel.out(\chi_{r13}).in(\chi_{r23}).E.rel.out(\chi_{r23}).in(\chi_{r32}).N.rel.out(\chi_{r32}),\chi_{r13}.\chi_{r23}.\chi_{r32} \\
 in(\chi_{r11}).E.S.rel.out(\chi_{r11}).in(\chi_{r22}).N.err\_out(\chi_{r22}).in(\chi_{r23}).E.rel.out(\chi_{r23}).in(\chi_{r32}).N.rel.out(\chi_{r32}),\chi_{r11}.\chi_{r22}.\chi_{r23}.\chi_{r32} \\ 
\end{array}
\]
}

This set produces the following sequences of positions: $ p_0 p_4 p_3 p_3 p_4 p_7 p_7 p_8 p_8$, $p_0 p_4 p_7 p_7 p_8 p_8 $, and also 
$ p_0 p_4 p_4 p_7 p_7 p_8 p_8$ and 
$p_0 p_4 p_3 p_3 p_4 p_4 p_7 p_7 p_8 p_8$. 
Note that the last trace is faulty (e.g.~ traces contain error actions $err\_out$, see below) since it contains a forced release $ err\_out(\chi_{2,2}) $ (see below). 
Consequently, the system \textit{does not respect} the policy $ \varphi_{IT} $ for the item. 
In particular, there are three faulty traces found by the analysis, which have the following common prefix:
{
\[
\begin{array}{l}
 in(\chi_{r11}).E.S.rel.out(\chi_{r11}).in(\chi_{r22}).N.out\_err(\chi_{r22}),\chi_{r11}.\chi_{r22} \\ 
\end{array}
\]
}
The reason is that the robot $ R_{2,2} $ is forced to release the item when attempting to move it into the bad position $ p_5 $.
Moreover, there is no faulty trace of actions over sensors, which means the system \textit{respects} the policies $ \varphi_{i,j} $ for sensors and therefore complies with it.



\end{example}


The analysis provides us with an approximation of the overall behaviour of the analysed process. 
Moreover, it is proved to be correct: 
the analysis indeed respects the operational semantics of G-Local $ \pi $-calculus, as shown by the following subject reduction result.

\begin{theorem} \textbf{(Subject Reduction)} 
$ (\rho,\kappa,\Gamma,\psi) \models^{\delta} P $ and $ P \xrightarrow{\mu}^* P' $, then $ (\rho,\kappa,\Gamma,\psi) \models^{\delta} P' $.
\end{theorem}

We can further prove that there always exists a 
a least choice of $(\rho,\kappa,\Gamma,\psi) $ that is acceptable for CFA rules, and therefore it always exists a least estimate.
This depends from the fact that the set of analysis estimates constitutes a Moore family.


\begin{theorem} \textbf{(Existence of estimates)} For all $ \delta, P $, the set
$
\{ (\rho,\kappa,\Gamma,\psi) | (\rho,\kappa,\Gamma,\psi) \models^{\delta} P \}
$
is a Moore family.
\end{theorem}

Moreover, our analysis offers information on the resource usage, included bad usages.
The component $\Gamma$ is indeed in charge of recording all the possible usage traces on each resource $r$. Actually, for each $r$, traces are composed of pairs 
$[(\phi, \eta), S\chi]$, where $S$ is made of  labels of the processes that acquired the resource $r$ and $\eta$ records every action on $r$, included the special actions
$in(\chi)$, $out(\chi)$ and $err\_out(\chi)$, that indicate that the process labelled $\chi$ may acquire and release (or it may be forced to release) the resource.
This information offers a basis for studying dynamic properties,
by suitably handling the safe over-approximation the CFA introduces. 
We want to focus now on the traces including special error actions, that we call {\em faulty}.
\begin{definition}
A trace $\eta \in \hat{\mathcal{A}}^*$ is {\em faulty} if it includes
$err\_out(\chi)$ for some $\chi \in \mathcal{L}$.
\end{definition}

In particular, on the one hand 
if the analysis contains faulty traces, then there is the {\em possibility} of policy violations,
while if all the traces are not faulty, then we can prove that 
policy violations cannot occur at run time, and therefore
that the processes correctly use their resources.

We can show it formally, as follows.
\begin{definition}
The process $P$, where $r$ is declared with policy $\phi$, $P$ {\em complies with} $\varphi$ for $r$, if and only if
$P \xrightarrow{\mu}^{*} P'$ implies that there is no $P''$ such that $P' 
\inter{\overline{\alpha(r)}} P''$, where $•\xrightarrow{\mu}^{*}$ is the reflexive and transitive closure of $\xrightarrow{\mu}$. 
\end{definition}

\begin{definition}
A process $P$, where $r$ is declared with policy $\varphi$, is said to {\em respect} $\varphi$ for $r$, if and only if
$$\exists(\rho,\kappa,\Gamma,\Psi).(\rho,\kappa,\Gamma,\psi)^{[\epsilon,\epsilon]}{P}   \mbox { and $\forall [(\varphi,\eta),S] \in \Gamma(r). \eta$ is not faulty }$$
\end{definition}


\begin{theorem}
If $P$ {\em respects} the policy $\varphi$ for $r$ then,
$P$ {\em complies with} $\varphi$.
\end{theorem}

\section{Concluding Remarks}
Our work combines the name-passing of the $\pi$-calculus with the publish-subscribe paradigm to cope with resource-awareness. We have shown that this has lead to a name passing process calculus with primitives for acquiring and releasing stateful resources.
Our research program is to provide formal mechanisms underlying the definition of a resource-aware programming model. The work reported in this paper provides a first step in this direction.
There is a number of ways in which our calculus could be extended. In terms of calculus design, we assumed a monadic request primitive for managing resource binding. This is a reasonable assumption for several cases. An interesting issue for future research is to extend the calculus with a polyadic request primitives asking for a finite number of resources. In terms of reasoning mechanisms, it would be interesting to exploit CFA techniques to develop methodologies to analyze the code in order to avoid bad accesses to resources. Also it would be interesting to apply the typing techniques  (behavioral types)  introduced in \cite{BDFZ09} to capture a notion of resource contract.

\bibliographystyle{eptcs}
\bibliography{references}

\end{document}